\documentclass[12pt]{article}

\usepackage{sbc-template}

\usepackage{graphicx,url}
\usepackage{float}
\usepackage{amsmath}
\usepackage{array}
\usepackage{booktabs}

\usepackage[brazil]{babel}
\usepackage[utf8]{inputenc}
\usepackage{hyperref}
\hypersetup{
  setpagesize  = false,
  colorlinks   = true,
  urlcolor     = blue,
  linkcolor    = black,
  citecolor    = black
}
\usepackage{setspace}

\usepackage[brazilian,hyperpageref]{backref}
\usepackage[alf]{abntex2cite}

\title{Empirical Evaluation of SMOTE in Android Malware Detection with Machine Learning: Challenges and Performance in CICMalDroid 2020}

\author{André Augusto Bortoli\inst{1}, Diego Ferreira Duarte\inst{1}, \\
Mariana Recamonde Mendoza Guerreiro\inst{1} }

\address{Instituto de Informática -- Universidade Federal do Rio Grande do Sul
  (UFRGS)\\
  Caixa Postal 15.064 -- 91.501-970 -- Porto Alegre -- RS -- Brazil
  \email{\{andre.bortoli,diego.duarte,mrmendoza\}@inf.ufrgs.br}
}

\begin{document} 

\maketitle

\begin{abstract}
Malware, malicious software designed to damage computer systems and perpetrate scams, is proliferating at an alarming rate, with thousands of new threats emerging daily. Android devices, prevalent in smartphones, smartwatches, tablets, and IoTs, represent a vast attack surface, making malware detection crucial. Although advanced analysis techniques exist, Machine Learning (ML) emerges as a promising tool to automate and accelerate the discovery of these threats. This work tests ML algorithms in detecting malicious code from dynamic execution characteristics. For this purpose, the CICMalDroid2020 dataset, composed of dynamically obtained Android malware behavior samples, was used with the algorithms XGBoost, Naïve Bayes (NB), Support Vector Classifier (SVC), and Random Forest (RF). The study focused on empirically evaluating the impact of the SMOTE technique, used to mitigate class imbalance in the data, on the performance of these models. The results indicate that, in 75\% of the tested configurations, the application of SMOTE led to performance degradation or only marginal improvements, with an average loss of 6.14 percentage points. Tree-based algorithms, such as XGBoost and Random Forest, consistently outperformed the others, achieving weighted recall above 94\%. It is inferred that SMOTE, although widely used, did not prove beneficial for Android malware detection in the CICMalDroid2020 dataset, possibly due to the complexity and sparsity of dynamic characteristics or the nature of malicious relationships. This work highlights the robustness of tree-ensemble models, such as XGBoost, and suggests that algorithmic data balancing approaches may be more effective than generating synthetic instances in certain cybersecurity scenarios.
\end{abstract}

\subsection{Use of Artificial Intelligence tools in work development}

In order to maintain a commitment to academic ethics and machine learning guidelines, the authors report that Google's natural language processing and text generation tool, Gemini, was used exclusively for text evaluation and concordance suggestions in order to improve the writing of this work. In addition, during writing in the Overleaf tool, its native integration with language processing models, Writefull, was also used for a similar purpose.

\section{Introdução}

According to \citeonline{Alsmadi2021}, malware (malicious software) is a generic term used to describe a program or code designed to harm a computer, network or server. Cybercriminals develop malware to discreetly infiltrate a computer system in order to violate or destroy sensitive data and computing environments. According to statistics from \citeonline{AVG}, in 2024, more than 300,000 new types of cyber threats were developed every day.

Due to the increasing sophistication and quantity of malicious codes, it is crucial to perform malware detection analysis. Many researchers strive to mitigate cyberattacks through various approaches \cite{Hadiprakoso2020}.

To perform malware analysis there are 3 main types of analysis:

\begin{itemize}
    \item \textbf{Static} - When the binary is analyzed in text mode, and is analyzed without executing it, the main technique used is \textit{string} analysis, where we look for text snippets that can indicate the behavior that the \textit{malware} executes;
    \item \textbf{Dynamic} - The \textit{malware} is executed in a controlled environment, with tools that help identify its behavior;
    \item \textbf{Reverse engineering or code analysis} - Tools are used to obtain the source code of the \textit{malware} and the analysis is done at the programming code level.
\end{itemize}

These analyses must be carried out carefully and by people with advanced knowledge of operating systems, networks, programming, and information security. As an alternative to the analyses above, resources can be used to automate threat recognition; Machine Learning (ML) can contribute to this task.

According to \citeonline{Soori2023}, ML is a branch of Artificial Intelligence (AI) focused on enabling computers and machines to imitate the way humans learn, perform tasks autonomously, and improve their performance and accuracy through experience and exposure to more data.

Smartphones, tablets and other IoT devices represent a large portion of the devices that access the internet. In 2024, the number of connected IoT devices is expected to grow by 13\%, reaching 18.8 billion globally \cite{Sinha2024}. Smartphones and tablets represent 94.2\% of all devices used to access the internet \cite{Howarth:5}. Among these, Android is the main operating system used.

In this context, it is justified to develop a work applying ML techniques to the CICMalDroid2020 dataset, a base with 11,598 samples of executions of malware developed for Android, with the machine learning algorithms: Extreme Gradient Boosting algorithm (XGBoost), Naïve Bayes (NB), Support Vector Classifier (SVC) and Random Forest (RF). To demonstrate how AI can contribute to the analysis of malware.

\section{Referencial Teórico}

\subsection{Técnicas para classificação de Malware em Android}

The classification and detection of malicious code on Android devices has been a topic of frequent study, with the application of a variety of features and ML algorithms being applied to provide defense tools for the mobile device technology landscape \cite{liu2022, muzaffar2022}. The features used in these analyses can be classified as static, dynamic and hybrid \cite{meijin2022, tanvirul2024}.

Static features refer to the direct extraction of Android Application Package (APK) files; for example, analysis of the \texttt{AndroidManifest.xml} file can reveal application components, required permissions, and hardware and software requirements \cite{park2018}. In parallel, dynamic features are obtained by running the application in a controlled environment, such as a sandbox, and monitoring its behavior; this type of analysis involves the observability of runtime activities, such as system calls, network traffic, data transmission, and resource consumption citesihag2021; the functionalities derived from system calls and binders\footnote{In the context of Android, a binder is an inter-process communication (IPC) mechanism that allows different components of an application or even distinct applications to exchange data and call each other's methods}, as used in the CICMalDroid2020 dataset (see subsection \ref{sec:maldroidds}), are examples of dynamic features analyzed in this scope.

A wide range of ML algorithms are applied to this problem, including Decision Tree (DT), RF, Support Vector Machines (SVM), NB, and Logistic Regression (LR) \cite{yuan2014, alkahtani2022, csahin2022, singh2021}. Deep learning models, including Convolutional Neural Networks (CNNs) and Recurrent Neural Networks (RNNs), have also been actively explored due to their potential in capturing complex patterns in data for malicious execution analysis. The suitability of each model depends on the nature, volume and dimensionality of specific features that are analyzed \cite{mahesh2019}.

\subsection{Base de dados: CICMalDroid 2020}
\label{sec:maldroidds}

\begin{figure}[H]
\centering
\includegraphics[width=\textwidth]{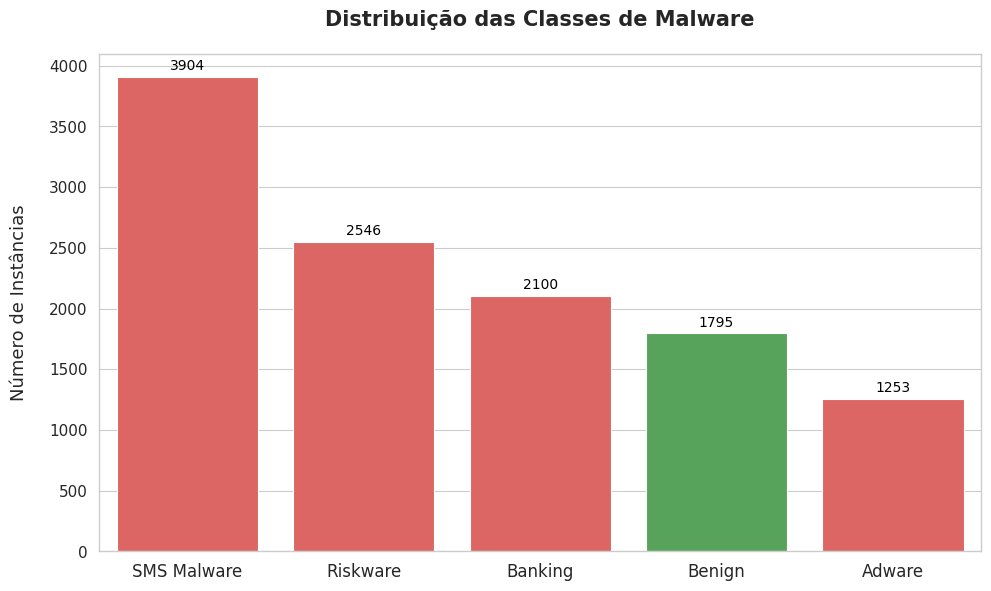}
\caption{Distribution of classes in the CICMalDroid 2020 database, the malicious categories colored in red and the benign category in green.}
\label{fig:distribuicao-classes}
\end{figure}

The CIC-MalDroid 2020 dataset, created at the Canadian Cyber Security Institute by \citeonline{mahdavifar2020}, contains 11,598 instances collected between 2017 and 2018 from sources that included VirusTotal\footnote{Virustotal, available at \url{https://www.virustotal.com} (accessed on 05/06/2025), is an \textit{online} service intended for analyzing files and URLs for viruses, \textit{worms}, \textit{trojans}, and other types of malicious content using antivirus and \textit{website scanners}}, Contagio\footnote{Contagio, available at \url{https://contagiodump.blogspot.com} (accessed on 05/06/2025), is a \textit{blog} that disseminates information on attacks and campaigns focused on cybersecurity} and AMD\footnote{Database created by \citeonline{fengguo2017} in an effort to replace the previous dataset, which was already outdated}. This database is divided into five categories, illustrated in Figure \ref{fig:distribuicao-classes}: "\textit{Adware}" (1,253 instances), "\textit{Banking}" (2,100 instances), "\textit{Riskware}" (2,546 instances), "\textit{SMS malware}" (3,904 instances) and "\textit{Benign}" (1,795 instances), with the "\textit{Benign}" category being equivalent to executions that do not present a risk to the user, that is, they do not execute malicious code. The choice of this database is justified by its recent nature, high volume, robustness of the collected characteristics and academic relevance \cite{zhao2018android, Hadiprakoso2020, shafin2021}.

\subsection{Algoritmos de ML}

Choosing an appropriate classification algorithm for a particular problem is a task that requires practice and expertise, each algorithm has its peculiarities and is based on certain assumptions \cite{raschka2019python}, for this reason, 4 algorithms from the Scikit-learn library\footnote{Scikit-learn, available at \url{https://scikit-learn.org/} (accessed on 06/05/2025), it is an open source library for the \textit{Python} programming language, developed specifically to facilitate the practical application of machine learning \cite{scikit-learn}} were selected, and carry out experiments to obtain conclusions from their results.

RF is a nonparametric approach that can accommodate different types of responses, such as categorical or quantitative outcomes. Furthermore, it can work with predictors of various scales or distributions and is suitable for applications in high-dimensional environments, where the number of predictors may be larger than the number of observations \cite{Jianchang2023}.

One machine learning approach to classification problems is Linear Support Vector Classifier (SVC) which works especially well with high-dimensional datasets such as text. A variant of the Support Vector Machine (SVM), Linear SVC finds the best hyperplane to split classes with the largest margin using a linear kernel \cite{Ray2025}.

XGBoost is a machine learning algorithm based on decision tree boosting, with the advantages of high flexibility, strong predictability, strong generalization ability, high scalability, high model training efficiency, and strong robustness \cite{Zhang2022}.

\citeonline{Chen2020} describes NB as an algorithm that treats each attribute equally. However, in real-world applications, different attributes play different roles in class discrimination, with some attributes being more important than others. For this reason, it is called \textit{Naïve}. Based on Bayes' Theorem, it works as a probabilistic classifier.

\subsection{Pré-processamento de Dados para ML}

Data preprocessing in ML involves preparing the raw data set available in a database for analysis and model training. This includes handling missing data, normalizing variables, reducing noise, and transforming the data into formats more suitable for specific algorithms. In the absence of the preprocessing step, the model loses generalization capacity and performance.

\subsubsection{Escalonamento de características}

Feature scaling is a normalization strategy that standardizes the sample space of variables in a database \cite{singh2020}. The absence of this process has a high potential impact on model performance, introducing problems for data generalization and biases in predictions. Three scaling approaches widely used in the literature are described below.

Standard normalization (also described as \textit{Z-score normalization}) transforms attributes to ensure that the mean is zero and the standard deviation is one. This transformation follows the formula $x_{\text{scaled} = \frac{x - \mu}{\sigma}}$ where $\mu$ represents the column mean and $\sigma$ its standard deviation. This method assumes a Gaussian distribution and is highly sensitive to outliers \cite{patro2015}.

Additionally, Min-Max normalization is limited to rescaling the values of the variable, typically using the scale [0, 1], consequently, the original distribution and probabilistic form are maintained. The transformation applied in this method is defined by the equation $x_{scaled} = \frac{x - X_{min}}{X_{max} - X_{min}}$. This approach is applicable when the data distribution is uniform and the minimum and maximum values are representative of the type of data analyzed by the model \cite{patro2015}.

Finally, robust normalization uses median and interquartile range for data scaling, which makes this method resistant to outliers. The equation that defines this method is $x_{scaled} = \frac{x - Q_2}{Q_3 - Q_1}$ where $Q_{2}$ is the median value and $Q_3 - Q_1$ represents the interquartile range \cite{brownlee2020}.

\subsubsection{Variable Selection}

Variable selection aims to address a problem known as the curse of dimensionality\footnote{The curse of dimensionality refers to the fact that the more attributes a dataset has, the more difficult it becomes to analyze, process, and model them, both for human analysts and for algorithms that identify patterns \cite{altman2018}} by identifying and retaining the most important features in the database and eliminating redundant or irrelevant ones \cite{guyon2003}. This process reduces computational costs both in the training stage and in the subsequent prediction of the model, contributes to the interpretation of the model, and has potential in the resulting generalization performance.

In this context, variance-based selection represents an algorithm known in the literature that removes columns with low variance, that is, constant or nearly constant columns, based on the premise that these tend to provide little or no information about the target variable.

Additionally, statistical analysis-based selection is a method that tests the individual relationship between possible classes in the target column and a given column. This correlation is commonly calculated using one of the following approaches: (i) The F-test (also described as \textit{ANOVA F-value}) measures the linear dependence between the variables and the target. This method is suitable for classification tasks or when analyzing columns with data that follow linear correlations; (ii) On the other hand, the Chi-square test evaluates the independence between categorical columns and the target column, and is suitable for discrete feature analysis.

\subsubsection{Dimensionality Reduction}

Similar to variable selection, dimensionality reduction aims to transform sets with many variables into less costly bases. However, unlike the selection process, dimensionality reduction applies transformations to the base in an attempt to preserve inherent information and patterns \cite{van2009}, essentially merging different parameters into new unified columns through a projection of the original values.

Among the approaches explored in this context, a recurrent one is Principal Component Analysis (PCA), which applies an orthogonal transformation to unify correlated variables into principal components while minimizing the correlation of the resulting components. As a consequence, the transformation applied by this method maximizes the variance in the design space, but also makes it dependent on normalization of the data set prior to its application.

In parallel, Truncated Singular Value Decomposition (SVD) works by decomposing the data presented by the original matrix into three matrices. Since it does not depend on a data centralization process, Truncated SVD, unlike PCA, can be applied to sparse data and does not require prior normalization for its calculation, making it useful in scenarios in which the original scale or sparsity of the data are important.

\subsection{Strategies for mitigating imbalanced data in ML}

Imbalanced class representation is a recurring problem in several domains in which ML techniques are applied and is characterized by the significant dominance of one class over the others, which can lead to models biased towards the target with numerical superiority \cite{he2009}. This phenomenon affects the cybersecurity domain and introduces problems for categorizing malicious activities since they often constitute a low fraction in relation to the total number of incidents.

\subsubsection{Resampling Techniques}

\textit{Oversampling} strategies increase the representation of minority classes by generating synthetic instances. The Synthetic Minority Over-sampling Technique (SMOTE) creates new examples by interpolating existing instances with the K-Nearest Neighbors (KNN) algorithm \cite{chawla2002}. This approach generates synthetic examples with greater diversity than alternatives such as simple duplication, reducing the risk of overfitting.

In parallel, undersampling strategies eliminate instances of the majority class to balance the class distribution. Although efficient, this approach presents risks related to the loss of useful information for generalization due to the potential reduction in the representativeness of the database.

\subsubsection{Algorithmic Approaches}

\textit{Ensemble} algorithms can mitigate imbalanced data by combining multiple models using the "wisdom of the crowd" strategy. In particular, \textit{gradient boosting} approaches iteratively focus on misclassified instances, effectively providing more attention to minority classes during training.

In addition, cost-sensitive algorithms incorporate different costs for miscategorization, penalizing minority class errors more severely than majority class errors.

\subsection{Hyperparameter Optimization}

Hyperparameter Optimization (HPO) is a process that systematically searches for optimal settings of an algorithm in order to maximize the performance of a resulting model \cite{bergstra2012}.

\subsubsection{Bayesian Optimization}

HPO strategies treat hyperparameter tuning as a black-box problem. This means that the algorithm makes no assumptions about the exact function that computes the model’s performance metric. This approach is necessary because finding the optimal parameters is inherently opaque, influenced by factors such as the variability of data preprocessing steps, the specific characteristics of the dataset, and the training algorithm employed. Consequently, evaluating any set of hyperparameters requires a complete cycle of model training and validation, which generates a demand for algorithms and techniques capable of efficiently approximating sufficiently optimal results.

\citeonline{shahriari2016} review how Bayesian optimization approaches address this problem. They do so by constructing probabilistic surrogate models of the objective function. These models allow the search process to be guided more intelligently, balancing the exploration of uncertain regions of the hyperparameter space with the validation of more promising configurations.

Among existing Bayesian approaches, the Tree-structured Parzen Estimator (TPE) algorithm models the conditional probability of hyperparameters given the reported performance across iterations, separating observations into high- and low-performance groups \cite{bergstra2012}. TPE estimates probability densities for each group and then selects parameters that maximize the ratio between the high- and low-performance probabilities. This approach has been successfully implemented in modern frameworks such as Optuna, which expands on the original TPE concept.

\subsubsection{Optuna}

Optuna is an HPO framework that implements Bayesian hyperparameter optimization techniques with improvements aimed at flexibility in use for production models, for example, the ability to define conditional and dynamic searches in the parameter space \cite{akiba2019}. In addition, Optuna incorporates pruning mechanisms to stop testing on unpromising models, native horizontal scalability, and visualization tools for evaluating and analyzing the relationships between hyperparameters and performance during training.

\subsubsection{Pruning mechanisms}

Pruning mechanisms are algorithms that determine the early termination of an iteration based on criteria that determine poor performance of the model being produced, which, when applied to the scale of several models created and evaluated in the HPO process, results in a substantial reduction in computational costs with low chances of impacting the quality of the resulting model.

Among the known methods, median-based pruning compares intermediate results with historical medians in equivalent steps, terminating early those models that do not meet this minimum requirement.

\subsubsection{Cross-Validation}
\label{sec:crossvalidation}

Cross-validation provides a way to measure the results of a model across partitions of the data set that are iteratively split into "folds" and assigned to the training or testing set \cite{stone1978}.

\subsubsection{Stratified Cross-Validation}

Stratified cross-validation maintains the proportion of classes across all folds, aiming to ensure a representative proportion of the data across all validation iterations and contributing to the mitigation of problems related to the generalization of the resulting models \cite{kohavi1995}.

\subsubsection{Nested Cross-Validation}

Nested cross-validation separates sections of the model using two iteration structures: an inner and an outer one \cite{varma2006}. Internal iteration is used to optimize hyperparameters and, after identifying the most promising model, its validation is performed using test data from the external iteration. By applying this approach, data contamination between the hyperparameter optimization process and the model evaluation phase is avoided.

\section{Methodology}

\subsection{Test environment}

All tests were collected in an environment provided by the Google Colab tool and are available and documented in Jupyter format (\texttt{.ipynb}) at \url{https://zenodo.org/records/15620300}. The execution type used adhered to the hardware accelerator with A100 GPU.

\subsection{Preprocessing Pipeline}

Preprocessing was implemented through the pipeline provided by the Scikit-learn library, allowing the automated chaining of transformations. Each model configuration included specific preprocessing steps adapted to its algorithmic needs: (i) for sanitizing unwanted features, variance-based selection was used; (ii) standard normalization strategies were used for normalization — applied to SVM algorithms, given their high sensitivity to differences in magnitude of variables —, Min-Max normalization and robust normalization; (iii) for dimensionality reduction, PCA and Truncated SVD were used; (iv) for variable selection, the F-test and Chi-square criteria were alternated.

\subsection{Tested Model Configurations}

24 different configurations were evaluated, grouped by algorithm category.

\subsubsection{NBs}

BernoulliNB (Chi-square Selection) (Aggressive): Assumes binary features with statistical feature selection (Chi-square). \texttt{k} was aggressively restricted to the range between 10 and 50.

BernoulliNB (Chi-square Selection): Similar to aggressive, but with \texttt{k} optimized in a wider space, from 200 to the total number of columns.

GaussianNB (PCA + Robust Scaling) (Aggressive): Uses outlier-resistant preprocessing (RobustScaler) with PCA dimensionality reduction. \texttt{n\_components} was aggressively optimized over the range 20 to 100.

GaussianNB (PCA + Robust Scaling): Same as aggressive, but with \texttt{n\_components} optimized over a wider space, from 50 to the total number of columns.

GaussianNB (F-test Selection) (Aggressive): Assumes Gaussian feature distributions after standard scaling. \texttt{k} was aggressively optimized over the range 10 to 50.

GaussianNB (F-test Selection): Similar to aggressive, but with \texttt{k} optimized over a wider space, from 50 to the total number of columns.

MultinomialNB (Chi-square + MinMax Selection) (Aggressive): Ensures non-negative features, a requirement of the multinomial assumption, with optimization of the \texttt{alpha} and \texttt{fit\_prior} parameters. A variable selection step using chi-square was also included with the \texttt{k} parameter being aggressively optimized in the range between 20 and 100

MultinomialNB (MinMax): Adheres to the previous standard, but without the variable selection step.

\subsubsection{SVMs}

LinearSVC (PCA) (Aggressive): Uses a linear kernel with standard scaling and dimensionality reduction by PCA. The regularization parameter C was optimized on a logarithmic scale in the range [0.1,100]. The dimensionality reduction, \texttt{n\_components}, was aggressively restricted to the range of 10 to 100 components, aiming for a lower dimensionality scenario

LinearSVC (PCA): Similar to the aggressive setting, but with the PCA parameter \texttt{n\_components} optimized on a wider search space, from 50 to the total number of columns.

LinearSVC (F-test + PCA selection) (Aggressive): Employs the linear kernel with standard scaling, variable selection using \texttt{SelectKBest} (F-test), and dimensionality reduction by PCA. The regularization parameter C was optimized on a logarithmic scale over the range [0.1, 100]. \texttt{n\_components} was aggressively optimized in the range 10–100. Similarly, \texttt{k} was aggressively restricted to the range 10–50.

LinearSVC (F-test Selection + PCA): Keeps the same characteristics as aggressive RBF SVC, but with the parameters \texttt{k} (from texttt{SelectKBest}) and \texttt{n\_components} (from PCA) optimized in a wider space: from 100 to the total number of columns (for the \texttt{k} parameter) and from 50 to the total number of columns (for the \texttt{n\_components} parameter).

RBF SVC (PCA) (Aggressive): Employs a Radial Basis Function kernel with standard scaling and PCA. The parameters C and gamma were optimized together. Dimensionality reduction (\texttt{n\_components}) followed the aggressive approach, restricted to the range between 20 and 100.

SVC RBF (PCA): Maintains the same characteristics as aggressive SVC RBF, but with the \texttt{n\_components} of PCA optimized over a wider space, from 50 to the total number of columns.

\subsubsection{RFs}

RF (Simple): Applies no preprocessing, taking advantage of the intrinsic robustness of the RF algorithm with respect to feature scaling. Optimized hyperparameters include \texttt{n\_estimators}, \texttt{max\_depth}, \texttt{min\_samples\_leaf}, \texttt{min\_samples\_split}, and \texttt{bootstrap}.

RF (PCA) (Aggressive): Applies principal component transformation (PCA). \texttt{n\_components} has been aggressively optimized over the range of 20 to 100.

RF (PCA): Identical to aggressive, but with \texttt{n\_components} optimized over a wider space, from 50 to the total number of columns.

RF (Truncated SVD) (Aggressive): Uses preprocessing with TruncatedSVD to reduce computational complexity. \texttt{n\_components} has been aggressively optimized over the range of 20 to 100.

RF (Truncated SVD): Similar to aggressive, but with \texttt{n\_components} optimized over a wider space, from 50 to the total number of columns.

RF (Selection F-test) (Aggressive): Uses preprocessing with \texttt{SelectKBest} (F-test) to identify the most informative columns. The \texttt{k} parameter has been aggressively restricted to the range of 20 to 100.

RF (Selection F-test): Similar to aggressive, but with \texttt{k} optimized over a wider space, from 100 to the total number of columns.

\subsubsection{Gradient Boosting}

XGBoost (Simple): Default XGBoost configuration with default scaling and optimization of \texttt{n\_estimators}, \texttt{learning\_rate}, \texttt{max\_depth}, \texttt{subsample}, and \texttt{colsample\_bytree}. Does not apply additional feature reduction or selection.

XGBoost (F-test Selection) (Aggressive): Adds a statistical column selection step. \texttt{k} has been aggressively restricted to the range between 10 and 50.

XGBoost (F-test Selection): Similar to aggressive, but with \texttt{k} optimized over a broader space, from 100 to the total number of columns.

\subsection{Treatment of Class Imbalance}

In order to improve the performance of the resulting models, the SMOTE technique was applied to address class imbalance. Each model was trained in two conditions: with and without SMOTE, using the pipeline provided by the \texttt{Imbalanced-learn} library for integration into the cross-validation flow.

\subsection{Hyperparameter Optimization}

HPO was implemented using Optuna with the following specifications:

\begin{itemize}
    \item Sampler: Tree-structured Parzen Estimator (\texttt{TPESampler});
    \item Pruning strategy: Median-based pruning;
    \item Number of attempts: 50 per configuration;
    \item Direction: Maximization of performance measured by weighted recall.
\end{itemize}

Search spaces were defined empirically specifically for each algorithm, including logarithmic distributions for regularization parameters and categorical distributions for discrete parameters.

\subsection{Evaluation Protocol}

\subsubsection{Implementation of Nested Cross-Validation}

The model evaluation process was structured with nested cross-validation, aiming to ensure an analysis with minimal bias: (i) the external repetition used 5 folds using the stratified strategy in order to ensure that the proportion of classes was maintained in each division of the data, with the progressive random state (43, 44, 45, 46, 47) to introduce controlled variability; (ii) the internal repetition, for hyperparameter optimization, used 5 folds with random division (\texttt{test\_size}=0.2); (iii) the main optimization metric was weighted recall, chosen specifically to consider class imbalance and prioritize the detection of minority classes.

\subsubsection{Reproducibility Settings}

In order to ensure experimental reproducibility, a fixed random seed (value 42) was adopted for all stochastic components of the pipeline. Seed variations were implemented deterministically, incrementing the base value for each external iteration (42 + \texttt{<index>} + 1), ensuring controlled diversity between partitions.

Stratification was systematically applied to all data splits to preserve the original class distribution, especially critical given the imbalance of the dataset. However, due to the use of the \texttt{cuML} library for GPU acceleration in model training, some operations presented non-deterministic behavior inherent to GPU parallelism. To mitigate these effects, \texttt{n\_streams=1} was configured in the tree-based algorithms and specific seeds were applied to all parameterizable components, although minor variations may persist due to the architectural limitations of GPU parallel computing.

934 / 5.000
\section{Results and Discussions}

The comparative evaluation of ML models with and without SMOTE applied to the CICMalDroid 2020 database revealed a consistent pattern: The application of SMOTE resulted in performance degradation or marginal improvements in 75\% of the tested configurations. This finding challenges the idea of performance improvement in the application of synthetic instances for malware analysis, specifically applied to the CICMalDroid 2020 dataset.

\subsection{NB}

\begin{figure}[H]
\centering
\includegraphics[width=\textwidth]{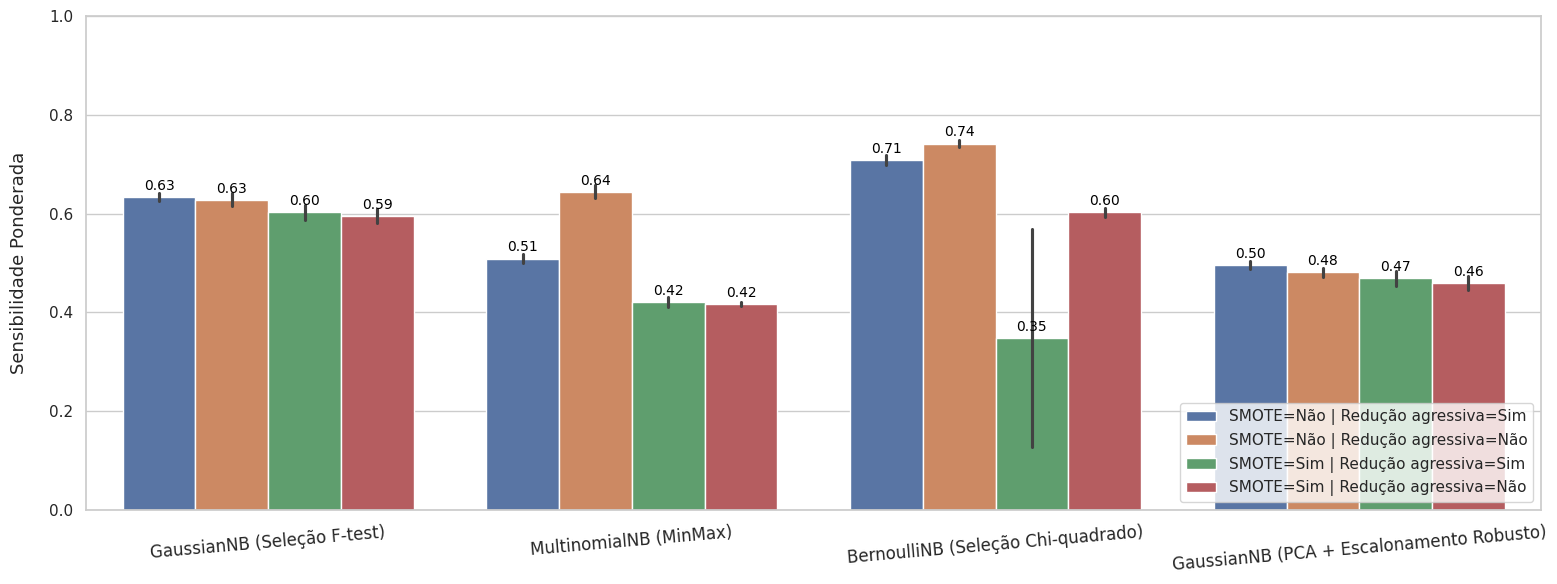}
\caption{Performance (weighted recall) by NB Models with and without SMOTE.}
\label{fig:nb-performance}
\end{figure}

NB algorithms obtained the worst results among all experiments, with SMOTE consistently degrading performance in all tested configurations, as detailed in Table \ref{tab:nb-results} and illustrated in Figure \ref{fig:nb-performance}, the assumption of natural independence of NB proved to be fundamentally incompatible with the complex relationships between variables, a characteristic inherent to malware detection.

\begin{table}[H]
\centering
\caption{Average performance (weighted recall) of NB models with and without SMOTE.}
\label{tab:nb-results}
\begin{tabular}{p{6cm}llc}
\toprule
\textbf{Model} & \textbf{Aggressive Reduction?} & \textbf{SMOTE?} & \textbf{Performance} \\
\midrule
BernoulliNB (Chi-square Selection)       & Não & Não & 74,18\% \\
\cline{3-4}
                                         &     & Sim & 70,87\% \\
\cline{2-4}
                                         & Sim & Não & 60,30\% \\
\cline{3-4}
                                         &     & Sim & 34,88\% \\
\midrule
GaussianNB (PCA + Robust Scaling) & Não & Não & 48,11\% \\
\cline{3-4}
                                         &     & Sim & 49,54\% \\
\cline{2-4}
                                         & Sim & Não & 45,93\% \\
\cline{3-4}
                                         &     & Sim & 46,88\% \\
\midrule
GaussianNB (F-test Selection)              & Não & Não & 62,82\% \\
\cline{3-4}
                                         &     & Sim & 63,43\% \\
\cline{2-4}
                                         & Sim & Não & 59,48\% \\
\cline{3-4}
                                         &     & Sim & 60,30\% \\
\midrule
MultinomialNB (MinMax)                   & Não & Não & 64,42\% \\
\cline{3-4}
                                         &     & Sim & 50,92\% \\
\cline{2-4}
                                         & Sim & Não & 41,72\% \\
\cline{3-4}
                                         &     & Sim & 42,13\% \\
\bottomrule
\end{tabular}
\end{table}

The most severe degradation occurred with the BernoulliNB variant using aggressive Chi-squared selection combined with SMOTE, dropping from an average of 60.30\% to 34.88\%. This drop reinforces the idea that the problem addressed is incompatible with the simplistic nature of the algorithm.

Notably, this same configuration exhibited high variance during training, suggesting instability when processing synthetic instances, indicating a possible failure to preserve the statistical properties of the original examples.

\subsection{SVM}

\begin{figure}[H]
\centering
\includegraphics[width=\textwidth]{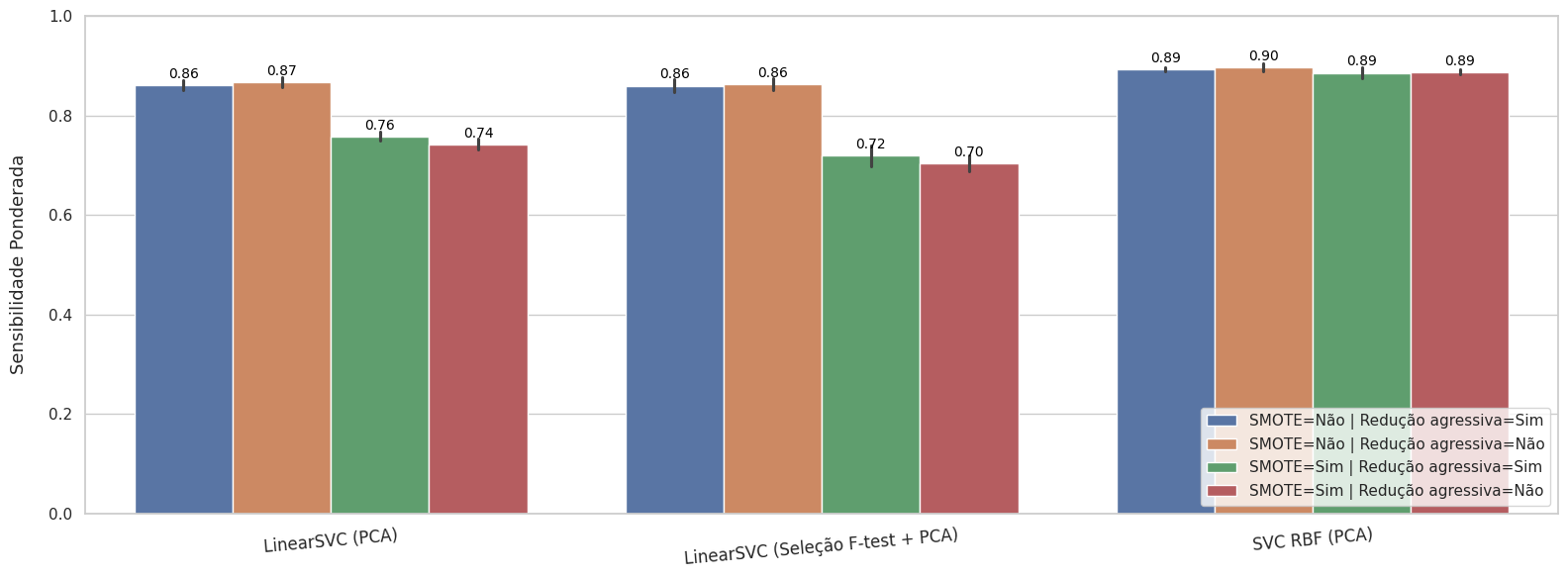}
\caption{Performance (weighted recall) by SVM Models with and without SMOTE.}
\label{fig:svm-performance}
\end{figure}

The SVM configurations demonstrated heterogeneous responses to the application of SMOTE, with an impact strongly related to the choice of \textit{kernel} used, as shown in Table \ref{tab:svm-results} and Figure \ref{fig:svm-performance}. This variance suggests evidence for nonlinear correlations between the features analyzed for malware analysis.

\begin{table}[H]
\centering
\caption{Average performance (weighted recall) of SVM models with and without SMOTE.}
\label{tab:svm-results}
\begin{tabular}{p{6cm}llc}
\toprule
\textbf{Model} & \textbf{Aggressive Reduction?} & \textbf{SMOTE?} & \textbf{Performance} \\
\midrule
LinearSVC (PCA)                  & Não & Não & 86,81\% \\
\cline{3-4}
                                 &     & Sim & 86,30\% \\
\cline{2-4}
                                 & Sim & Não & 74,13\% \\
\cline{3-4}
                                 &     & Sim & 75,69\% \\
\midrule
LinearSVC (Seleção F-test + PCA) & Não & Não & 86,06\% \\
\cline{3-4}
                                 &     & Sim & 86,37\% \\
\cline{2-4}
                                 & Sim & Não & 69,92\% \\
\cline{3-4}
                                 &     & Sim & 70,93\% \\
\midrule
SVC RBF (PCA)                    & Não & Não & 89,50\% \\
\cline{3-4}
                                 &     & Sim & 89,40\% \\
\cline{2-4}
                                 & Sim & Não & 89,45\% \\
\cline{3-4}
                                 &     & Sim & 88,76\% \\
\bottomrule
\end{tabular}
\end{table}

Linear SVM models showed little change when combined with SMOTE. LinearSVC with PCA suffered a minimal impact (0.51 percentage points decrease), while LinearSVC with feature selection and PCA showed a small improvement (0.31 percentage points). However, under aggressive dimensionality reduction, an interesting result occurred: SMOTE actually improved performance by 1.56 and 1.01 percentage points, respectively. This suggests that synthetic samples can help compensate for the loss of information during aggressive feature reduction.

SVMs with \textit{kernel} RBF maintained more stable performance, with degradation limited to 0.10-0.76 percentage points across all configurations. This stability indicates that the nonlinear decision boundaries created by \textit{kernels} RBF are more robust to the introduction of synthetic samples.

The differential impact based on \textit{kernel} type provides evidence that Android \textit{malware} detection requires non-linear modeling approaches, as linear \textit{kernels} consistently underperformed their RBF counterparts by 2-3 percentage points regardless of SMOTE application.

\subsection{RF}

\begin{figure}[H]
\centering
\includegraphics[width=\textwidth]{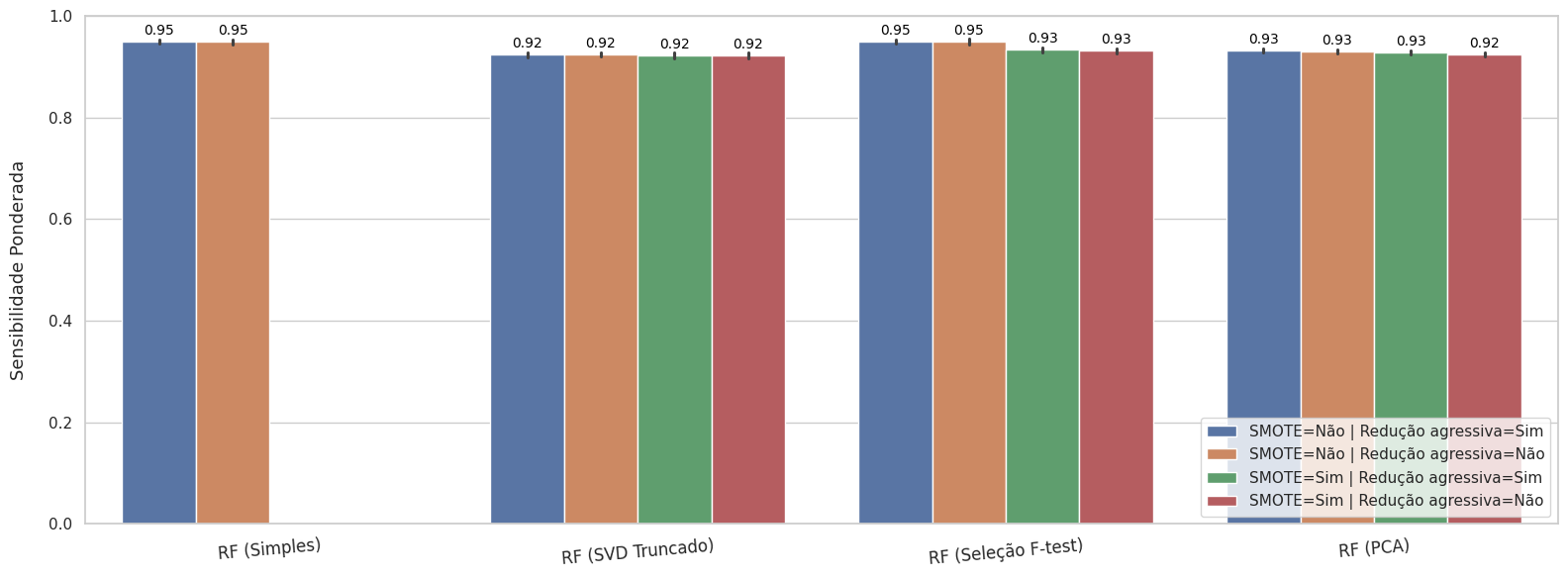}
\caption{Performance (weighted recall) by RF Models with and without SMOTE.}
\label{fig:rf-performance}
\end{figure}

Os modelos de RF demonstraram uma performance superior e consistente dentre as configuraçãoes, conforme detalhado na Tabela \ref{tab:rf-results} e apresentado na Figura \ref{fig:rf-performance}. Este algoritmo atingiu os resultados mais estáveis, com SMOTE produzindo impacto mínimo (tanto positivo quanto negativo) entre as diferentes abordagens testadas.

\begin{table}[H]
\centering
\caption{Average performance (weighted recall) of RF models with and without SMOTE.}
\label{tab:rf-results}
\begin{tabular}{p{6cm}llc}
\toprule
\textbf{Model} & \textbf{Aggressive Reduction?} & \textbf{SMOTE?} & \textbf{Performance} \\
\midrule
RandomForest (Simples)        & Não & Não & 94,86\% \\
\cline{3-4}
                              &     & Sim & 94,94\% \\
\midrule
RandomForest (PCA)            & Não & Não & 93,04\% \\
\cline{3-4}
                              &     & Sim & 93,18\% \\
\cline{2-4}
                              & Sim & Não & 92,41\% \\
\cline{3-4}
                              &     & Sim & 92,75\% \\
\midrule
RandomForest (SVD Truncado)   & Não & Não & 92,37\% \\
\cline{3-4}
                              &     & Sim & 92,31\% \\
\cline{2-4}
                              & Sim & Não & 92,24\% \\
\cline{3-4}
                              &     & Sim & 92,15\% \\
\midrule
RandomForest (Seleção F-test) & Não & Não & 94,88\% \\
\cline{3-4}
                              &     & Sim & 94,94\% \\
\cline{2-4}
                              & Sim & Não & 93,15\% \\
\cline{3-4}
                              &     & Sim & 93,28\% \\
\bottomrule
\end{tabular}
\end{table}

As configurações RF Simples e RF com Seleção de Atributos (não agressiva) alcançaram os mais altos níveis de desempenho (94,86\% e 94,87\%, respectivamente), com o SMOTE proporcionando melhorias marginais de 0,07 e 0,06 pontos percentuais. Embora essas melhorias sejam estatisticamente negligenciáveis, a estabilidade do desempenho do RF sugere uma robustez inerente ao desequilíbrio de classes que pode tornar a \textit{oversampling} sintética redundante.

Técnicas de redução de dimensionalidade (PCA, SVD) reduziram consistentemente o desempenho do RF, independentemente da aplicação do SMOTE, sugerindo que o mecanismo de seleção de atributos do \textit{ensemble} é mais eficaz do que abordagens baseadas em pré-processamento aplicadas a este conjunto de dados. A capacidade do algoritmo RF de lidar com espaços de alta dimensionalidade e a seleção implícita de atributos por meio de subamostragem aleatória parecem ser bem adequadas às características do conjunto de dados CICMalDroid 2020.

\subsection{XGBoost}

\begin{figure}[H]
\centering
\includegraphics[width=\textwidth]{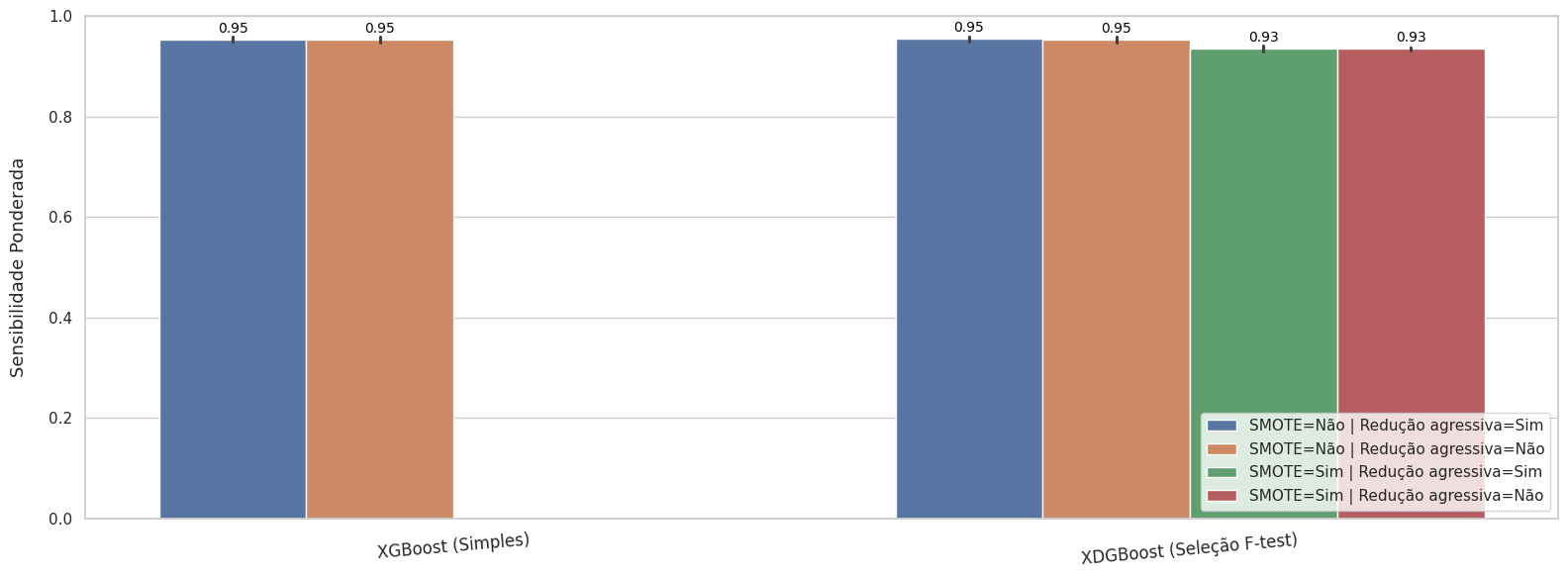}
\caption{Desempenho (Sensibilidade Ponderada) por Modelos XDGBoost com e sem SMOTE.}
\label{fig:xgboost-performance}
\end{figure}

Os algoritmos de XGBoost obtiveram melhores performances, com resultados minimamente superiores ao utilizar seleção de variáveis não agressiva e aplicação de SMOTE, conforme mostrado na Tabela \ref{tab:xgboost-results} e na Figura \ref{fig:xgboost-performance}.

\begin{table}[H]
\centering
\caption{Performance (weighted recall) by XGBoost Models with and without SMOTE.}
\label{tab:xgboost-results}
\begin{tabular}{p{6cm}llc}
\toprule
\textbf{Model} & \textbf{Aggressive Reduction?} & \textbf{SMOTE?} & \textbf{Performance} \\
\midrule
XGBoost (Simples)        & Não & Não & 95,29\% \\
\cline{3-4}
                          &     & Sim & 95,31\% \\
\midrule
XGBoost (Seleção F-test) & Não & Não & 95,32\% \\
\cline{3-4}
                          &     & Sim & 95,40\% \\
\cline{2-4}
                          & Sim & Não & 93,40\% \\
\cline{3-4}
                          &     & Sim & 93,46\% \\
\bottomrule
\end{tabular}
\end{table}

A configuração do XGBoost com Seleção de Atributos alcançou o desempenho mais alto (95,40\%) com a aplicação do SMOTE, representando uma melhoria de apenas 0,11 pontos percentuais em relação à alternativa simples sem SMOTE. Da mesma forma, a configuração XGBoost Simples não mostrou praticamente nenhuma alteração (95,31\% contra 95,29\%). Essas diferenças marginais estão dentro da variância típica do modelo e não podem ser consideradas melhorias estatisticamente significativas.

É importante notar que o XGBoost implementa estratégias algorítmicas nativas para lidar com conjuntos de dados desequilibrados através de sua estrutura de \textit{gradient boosting} e mecanismos internos de ponderação de classes. Essa capacidade inerente provavelmente explica o porquê da superamostragem sintética externa oferecer um benefício adicional mínimo — o algoritmo já otimiza para cenários de classificação desequilibrados.

O desempenho superior do XGBoost em comparação com outros algoritmos (95,40\% contra 94,94\% para RF, 89,45\% para SVM e 74,18\% para NB) estabelece o \textit{gradient boosting} como a abordagem mais eficaz para a detecção de \textit{malware} Android neste conjunto de dados. O processo de aprendizado sequencial do algoritmo e as técnicas de regularização parecem ser particularmente adequados às relações de atributos inerentes às aplicações de cibersegurança.

\section{Conclusão}

Dentre os 24 modelos testados, 75\% demonstraram degradação de performance ou benefícios marginais com a aplicação do SMOTE, sendo a performance média de perda equivalente a 6,14 pontos percentuais dentre os modelos degradados (2,85 removendo o valor atípico representado pelo modelo treinado com o algoritmo BernoulliNB). Apenas 25\% dos modelos apresentaram melhoras maiores do que 0,5\%, as quais somam uma média de aproximadamente 1,06 pontos percentuais.

A falha do SMOTE neste contexto pode ser atribuída a alguns fatores: (i) Visto que o SMOTE utiliza o algoritmo KNN para interpolação dos exemplos, há a possibilidade de que este não represente suficientemente as relações entre as colunas que caracterizam um comportamento malicioso; (ii) Dado que a base de dados CICMalDroid 2020 é bastante esparsa, o SMOTE pode ter problemas, dada sua natureza baseada em distância, em gerar exemplos suficientes, mesmo com a redução agressiva de dimensionalidade; (iii) As bordas de decisão entre maligno e benigno reais podem ser muito complexas para o algoritmo de interpolação proposto pelo SMOTE.

Por fim, infere-se, baseado nos resultados propostos, que algoritmos baseados em \textit{ensembles} de árvores (como RF e XGBoost) são mais robustos às características desta base de dados; modelos lineares foram particularmente afetados pela qualidade dos dados sintéticos que foram gerados; e a estratégia de balanceamento de dados algorítmica empregada pelo XGBoost foi mais eficiente do que o SMOTE para melhoria de resultados aplicados neste contexto.

\subsection{Trabalhos Futuros}

A partir do presente artigo, alguns caminhos podem ser traçados a fim de aprimorar a pesquisa destinada a melhorias na performance de algoritmos de detecção de \textit{malware}, especialmente relacionado a desbalanceamento de classes, dentre esses, os autores explicitam: (i) Investigação de técnicas alternativas para \textit{oversampling}; (ii) Investigação de técnicas mais elaboradas orientadas à engenharia manual de variáveis, buscando entender a correlação entre diferentes ações executadas por um programa e sua maliciosidade; (iii) Aplicação de métodos em \textit{ensemble} além dos algoritmos baseados em árvores; (iv) Investigação da performance e aplicabilidade de algoritmos de Aprendizado Profundo na detecção de \textit{malware} e/ou extração de características que possam ser analisadas por modelos menos custosos.

\bibliography{sbc-template}

\end{document}